\documentclass{article}
\begin{document}
\title{\large \bf  Nonsingular Increasing Gravitational Potential
for the Brane in 6D}
\author{{\bf Merab Gogberashvili$^a$ and Douglas Singleton$^b$} \vspace{0.3cm} \\
$^a$ Andronikashvili Institute of Physics, 6 Tamarashvili Str. \\
Tbilisi 0177, Georgia \\
\vspace{0.3cm}
{\sl E-mail: gogber@hotmail.com }\\
$^b$ Physics Dept., CSU Fresno, 2345 East San Ramon Ave.
M/S 37 \\
Fresno, CA 93740-8031, USA\\
{\sl E-mail: dougs@csufresno.edu} }
\maketitle
\begin{abstract}
\quotation{We present a new (1+3)-brane solution to Einstein
equations in (1+5)-space. As distinct from previous models this
solution is free of singularities in the full 6-dimensional
space-time. The gravitational potential transverse to the brane is
an increasing (but not exponentially) function and asymptotically
approaches a finite value. The solution localizes the zero modes
of all kinds of matter fields and Newtonian gravity on the brane.
An essential feature of the model is that different kind of matter
fields have different localization radii.}
\end{abstract}
\vskip 0.3cm
{ \sl PACS: 11.10.Kk, 04.50.+h, 98.80.Cq}
\vskip 0.5cm

The scenario where our world is associated with a brane embedded
in a higher dimensional space-time with non-factorizable geometry
has attracted a lot of interest since the appearance of papers
\cite{ADD,Go,RaSu}. In this model gravitons, which are allowed to
propagate in the bulk, are confined on the brane because of a
warped geometry. However, there are difficulties with the choice
of a natural trapping mechanism for some matter fields. For
example, in the existing (1+4)-dimensional models spin $0$ and
spin $2$ fields are localized on the brane with an exponentially
decreasing gravitational warp factor, spin $1/2$ field are
localized on the brane with an increasing factor \cite{BaGa}, and
spin $1$ fields are not localized at all \cite{Po}. For the case
of (1+5)-dimensions it was found that spin $0$, spin $1$ and spin
$2$ fields are localized on the brane with a decreasing warp
factor and spin $1/2$ fields again are localized with an
increasing factor \cite{Od}.

The reason why there are problems with localization of fermions
in warped geometries is that in the Lagrangians for the fermions
there appears an increasing exponential, coming from the metric
tensors with upper indices, $g^{AB}$, and from the tetrads,
$h^{\bar{A}}_B$. As a result the action integral over the extra
coordinates diverges, which is the signal for the
non-localization of the fermionic fields. In both (1+4)-, or
(1+5)-space models with warped geometry one is required to
introduce some non-gravitational interaction in order to localize
all the Standard Model particles.

For reasons of economy and to avoid charge universality
obstruction \cite{DuRuTi} one would like to have a universal
gravitational trapping mechanism for all fields. In our previous
papers we found such a solution of the 6-dimensional Einstein
equations in (2+4)- and (1+5)-spaces, which localized all kind of
bulk fields on the brane \cite{GoMi,GoSi}. These solutions
contain non-exponential scale factors, which increase from the
brane, and asymptotically approach a finite value at infinity. In
the paper \cite{Oda} the solution of \cite{GoSi} was generalized
to the case of $n$ dimensions. In this paper we present a new
solution to the Einstein equations in (1+5)-space, which, similar to
the models \cite{GoMi,GoSi}, also localizes all kind of physical
fields on the brane, but is free of singularities in the full
space-time. Because of this feature of the solution by setting realistic
boundary conditions we are able to fix the free parameters of the
model.

The general form of action of the gravitating system in six
dimensions is
\begin{equation} \label{action}
S = \int d^6x\sqrt{- ^6g}\left[\frac{M^4}{2}(^6R + 2 \Lambda) +
 ^6L \right] ~,
\end{equation}
where $\sqrt{-^6g}$ is the determinant, $M$ is the fundamental
scale, $^6R$ is the scalar curvature, $\Lambda$ is the
cosmological constant and $^6L$ is the Lagrangian of matter
fields. All of these quantities are six dimensional.

The 6-dimensional Einstein equations with stress-energy tensor
$T_{AB}$ are
\begin{equation} \label{Einstein6}
^6R_{AB} - \frac{1}{2} g_{AB}~^6R = \frac{1}{M^4}\left(\Lambda
g_{AB} + T_{AB}\right)~.
\end{equation}
Capital Latin indices run over $A, B,... = 0, 1, 2, 3, 5, 6 $.

As in the papers \cite{GoMi,GoSi} for the metric of the
6-dimensional space-time we choose the ansatz
\begin{equation}\label{ansatzA}
ds^2  = \phi ^2 (r)\eta _{\alpha \beta } (x^\nu  )dx^\alpha
dx^\beta   - \lambda (r)(dr^2  + r^2d\theta ^2)~ ,
\end{equation}
where the Greek indices $\alpha, \beta,... = 0, 1, 2, 3$ refer to
4-dimensional coordinates. The metric of ordinary 4-space,
$\eta_{\alpha \beta }(x^\nu)$, has the signature $(+,-,-,-)$. The
functions $\phi (r)$ and $\lambda (r)$ depend only on the extra
radial coordinate, $r$, and thus are cylindrically symmetric in
the transverse polar coordinates ($0 \le r < \infty$, $0 \le
\theta < 2\pi$).

The ansatz (\ref{ansatzA}) is different from the metric
investigated in (1+5)-space brane models with warped geometry
\cite{Od,Gr,GhSh}
\begin{equation}\label{ansatzB}
ds^2  = \phi ^2 (r)\eta _{\alpha \beta } (x^\nu  )dx^\alpha
dx^\beta   - dr^2  - \lambda (r)d\theta ^2~ .
\end{equation}
In (\ref{ansatzA}) the independent metric function of the extra
space, $\lambda (r)$, serves as a conformal factor for the
Euclidean 2-dimensional metric of the transverse space, just as
the function $\phi^2(r)$ does for the 4-dimensional part. However,
in (\ref{ansatzB}) the function $\lambda (r)$ multiples only the
angular part of the metric and corresponds to a cone-like geometry
of a string-like defect with a singularity on the brane at $r =
0$. Only in the trivial case $\lambda = r^2$ is the metric
(\ref{ansatzB}) regular on the brane.

The stress-energy tensor $T _{AB}$ is assumed to have the form
\begin{equation} \label{source}
T_{\mu\nu} = - g_{\mu\nu} F(r), ~~~~~ T_{ij} = - g_{ij}K(r),
     ~~~~~ T_{i\mu} = 0 ~.
\end{equation}
Using the ansatz (\ref{ansatzA}), the energy-momentum conservation
equation
\begin{equation}\label{energy-con}
\nabla^A T_{AB} = \frac{1}{\sqrt{-^6 g}}
\partial _A (\sqrt{-^6 g}T^{AB}) + \Gamma ^B _{CD} T^{CD} = 0~,
\end{equation}
gives a relationship between the two source functions $F(r)$ and
$K(r)$ from (\ref{source})
\begin{equation} \label{deltaT}
K^{\prime} + 4\frac{\phi^\prime}{\phi} \left(K - F \right) = 0 ~.
\end{equation}

We want to point out a problem associated with the source
functions (\ref{source}). In general the Einstein equations have
an infinite number of solutions generated by different matter
energy-momentum tensors, most of which have no clear physical
meaning. There is a great freedom in the choice of $F(r)$ and
$K(r)$; there are no other restrictions, except (\ref{deltaT}), on
their form. It is not easy to construct realistic source functions
from fundamental matter fields so that the brane is a stable,
localized object. We shall determine the sources $F(r)$ and $K(r)$
from some general physical assumptions that they are smooth
functions of the radial coordinate $r$, describe a continuous
matter distribution for all $r$, and that they decrease outside
the brane $r>\epsilon $, where $\epsilon$ is the brane width.

For a string-like defect, which corresponds to the metric ansatz
(\ref{ansatzB}), a set of $n$ scalar functions with a Higgs
potential (which breaks the global $SO(n)$ symmetry to $SO(n-1)$)
are used as a source to make the brane stable. In this case
topological arguments guarantee the stability of the brane because
the $(n-1)$ homotopy group is the integers: $\Pi_{n-1}(SO(n-1)) =
Z$ (see, for example \cite{Od}).

In our case we require the condition of classical stability
\cite{Go1}, that the total momentum of the brane-matter
configuration in the direction of the extra dimensions is zero
\begin{equation}  \label{P}
P_i=\int t_i^AdS_A = 0 ~,
\end{equation}
where $t_A^B$ is the total energy-momentum pseudo-tensor of
gravitation plus matter fields on the brane. For $t_A^B$ one can
choose, for example, the so-called Lorenz energy-momentum complex
\begin{equation}  \label{t}
t_A^B=\frac{1}{\sqrt{-^6g}}\partial_C\left[\sqrt{-^6g}
[g^{BD}g^{CE}(\partial _Dg_{AE}-\partial _Eg_{AD})]\right]~.
\end{equation}

To obey the stability condition (\ref{P}), brane solutions must
satisfy the conditions
\begin{equation} \label{ti}
t_i^\alpha = t_i^j = 0~.
\end{equation}
Using (\ref{t}) from (\ref{ti}) we obtain that the metric tensor
of the transverse space is a function of $r$ only and that
$g_{iA}=0$. These conditions are satisfied by both ans{\"a}tze
(\ref{ansatzA}) and (\ref{ansatzB}). One can have a condition
similar to (\ref{ti}) separately on the gravitational
energy-momentum pseudo-tensor, which gives \cite{Go1}
\begin{equation} \label{gGamma}
\partial_i\Gamma_{\beta \gamma }^\alpha = 0~,
\end{equation}
where $\Gamma_{\beta \gamma }^\alpha$ are components of the
6-dimensional Christoffel symbols. From (\ref{gGamma}) follows the
standard form of the 4-dimensional part of the brane metric
tensor, $g_{\alpha\beta} = \phi^2(r) \eta_{\alpha\beta}(x^\nu)$.
For the case of more then six dimensions  $\theta$ dependent
coefficients will appear in the angular part of the metrics
(\ref{ansatzA}) and (\ref{ansatzB}) and the condition of classical
stability (\ref{P}) fails. This may be one of the reasons why
brane solutions in more than six dimensions are singular
\cite{GhRoSh}.

The condition of classical stability (\ref{P}) is satisfied for
both metrics (\ref{ansatzA}) and (\ref{ansatzB}). It is expected
that, as for the string-like defect (\ref{ansatzB}), one can find
for (\ref{ansatzA}) particle physics models for the source
(\ref{source}) corresponding to stable brane configurations.

To solve equations (\ref{Einstein6}) we require that the
4-dimensional Einstein equations have the ordinary form without a
cosmological term
\begin{equation} \label{Einstein4}
R_{\mu\nu} - \frac{1}{2} \eta_{\mu\nu} R = 0~.
\end{equation}
The Ricci tensor in four dimensions $R_{\alpha\beta}$ is
constructed from the 4-dimensional metric tensor
$\eta_{\alpha\beta}(x^{\nu})$ in the standard way. Then with the
ans{\"a}tze (\ref{ansatzA}) and (\ref{source}) the Einstein field
equations (\ref{Einstein6}) become
\begin{eqnarray}\label{Einstein6a}
3 \frac{\phi^{\prime \prime}}{\phi} + 3 \frac{\phi ^{\prime}}{r
\phi} + 3 \frac{(\phi^{\prime})^2}{\phi ^2} +
\frac{1}{2}\frac{\lambda ^{\prime \prime}}{\lambda }
-\frac{1}{2}\frac{(\lambda^{\prime})^2}{\lambda^2} +
\frac{1}{2}\frac{\lambda^{\prime}}{r\lambda }
= \frac{\lambda }{M^4}[F(r) - \Lambda ] ~, \nonumber \\
\frac{\phi^{\prime} \lambda^{\prime}}{\phi \lambda } + 2
\frac{\phi^{\prime}}{r\phi} + 3 \frac{(\phi^{\prime})^2}{\phi
^2} = \frac{\lambda }{2 M^4}[K(r) - \Lambda ] ~, \\
2\frac{\phi^{\prime \prime}}{\phi} - \frac{\phi^{\prime} \lambda
^{\prime}}{\phi \lambda } + 3\frac{(\phi^{\prime})^2}{\phi^2} =
\frac{\lambda }{2 M^4}[K(r) - \Lambda ] ~, \nonumber
\end{eqnarray}
where the prime $=\partial / \partial r$. These equations are for
the $\alpha \alpha$, $rr$, and $\theta \theta$ components
respectively.

Subtracting the $rr$ from the $\theta \theta$ equation and
multiplying by $\phi / \phi ^{\prime}$ we arrive at
\begin{equation}
\label{phi-g} \frac{\phi^{\prime \prime}}{\phi^{\prime}} -
\frac{\lambda^{\prime}}{\lambda } -\frac{1}{r} = 0 ~.
\end{equation}
This equation has the solution
\begin{equation}
\label{g} \lambda (r)= \frac{\rho^2 \phi ^{\prime}}{r} ~,
\end{equation}
where $\rho$ is an integration constant with units of length.

System (\ref{Einstein6a}), after the insertion of (\ref{g}),
reduces to only one independent equation. Taking either the $rr$,
or $\theta \theta$ component of these equations and multiplying it
by $r\phi^4$ gives
\begin{equation}\label{rr}
r \phi^3 \phi^{\prime \prime} + \phi^3 \phi^{\prime} + 3r\phi^2
(\phi^{\prime})^2 = \frac{\rho^2 \phi^4 \phi^{\prime}}{2 M^4}
[K(r) - \Lambda ] ~.
\end{equation}

In our previous paper \cite{GoSi} the source functions $F(r)$ and
$K(r)$ outside the core $r>\epsilon $ were taken to have the form
\begin{equation} \label{source1}
F(r >\epsilon ) = K(r>\epsilon) =  \frac{f}{\phi^2}~,
\end{equation}
where $f$ is some constant. Then from the Einstein equations the
following solution was found
\begin{equation} \label{phi1}
\phi = a \frac{r^b - c^b}{r^b + c^b} ~,
\end{equation}
where
\begin{equation} \label{c}
a = \sqrt{\frac{5f}{3\Lambda}}~, ~~~~~ b = \frac{a\Lambda
\rho^2}{5M^4}~, ~~~~~ c^b = \epsilon^b \frac{a - 1}{a + 1}
\end{equation}
are integration constants. The solution (\ref{phi1}) is an
increasing function from the brane to some finite value at
infinity
\begin{equation} \label{a}
\phi (\infty) = a = \sqrt{\frac{5f}{3\Lambda}} > 1 ~.
\end{equation}
The factor $1/\phi^2(r)$ has $\delta$-like behavior outside the
core and the source functions (\ref{source1}) decrease as
required.

In \cite{GoSi} it was shown that the solution (\ref{phi1})
provides a universal, gravitational trapping for all kinds of
matter fields. However, in this model we did not specify source
functions on the brane $0\le r \le \epsilon$ and there were a
large number of free parameters. Now we want to choose the source
functions $F(r)$ and $K(r)$ everywhere, so that the solution
$\phi$ will localize all kind of physical fields on the brane and
be a regular function in the full 6-dimensional space-time.

We require for $\phi$ the following boundary conditions near the
origin $r=0$
\begin{equation} \label{r=0}
\phi (r\rightarrow 0) \approx 1 + d r^2 ~, ~~~ \phi^\prime
(r\rightarrow 0) \approx 2 d r~,
\end{equation}
where $d$ is some constant. At infinity we want $\phi (r)$ to
behave as
\begin{equation} \label{asimptotics}
\phi (r \rightarrow \infty) \rightarrow a  ~,~~~~~\phi^\prime (r
\rightarrow \infty) \rightarrow 0 ~,
\end{equation}
where $a > 1$ is some constant. Since the function $\phi^\prime$
is proportional to the metric of the extra 2-space, the boundary
conditions (\ref{asimptotics}) imply that at infinity $\lambda
\rightarrow 0$ and the effective geometry is 4-dimensional.

The source functions $F(r)$ and $K(r)$, which satisfy restriction
(\ref{deltaT}) and give a desirable solution were found recently
in the paper \cite{Mi}
\begin{equation} \label{FK}
F(r) = \frac{f_1}{2 \phi^2} +\frac{3 f_2}{4 \phi}~, ~~~~~ K(r) =
\frac{f_1}{\phi^2} +\frac{f_2}{\phi}~,
\end{equation}
where $f_1 , f_2$ are constants. Note that these source functions
do not have a vanishing value at $r \rightarrow \infty$, due to
the asymptotic behavior of $\phi$ given in (\ref{asimptotics}).

Substituting (\ref{FK}) into (\ref{rr}), taking its first integral
and setting the integration constant to zero yields \cite{GoMi,Mi}
\begin{equation} \label{first}
r \phi^\prime = \frac{\rho ^2 \Lambda}{10 M^4} \left(\frac{5
f_1}{3 \Lambda} + \frac{5 f_2}{4 \Lambda}\phi - \phi^2 \right) ~.
\end{equation}
By introducing the parameters $A$ and $a$ such that
\begin{equation} \label{parameters}
\frac{\rho^2 \Lambda}{10 M^4} =A ~,~~~~~ f_1 = -\frac{3
\Lambda}{5} a~, ~~~~~ f_2=\frac{4 \Lambda}{5}(a+1)~,
\end{equation}
equation (\ref{first}) becomes
\begin{equation} \label{firsta}
r \phi^\prime = A [- a+(a+1) \phi -\phi^2 ] ~.
\end{equation}
Equation (\ref{firsta}) is easy to integrate \cite{Mi}
\begin{equation} \label{phi}
\phi = \frac{c^b+ a r^b}{c^b+ r^b}~,
\end{equation}
where $b = A(a-1)$ and $c$ are integration constants. From the
boundary conditions (\ref{r=0}) it follows
\begin{equation} \label{b}
b = A(a-1) = 2 ~.
\end{equation}

The width of the brane $\epsilon$ corresponds to the inflection
point of the function $\phi$ (where the second derivative of
$\phi$ become zero and then changes sign). From the condition
$\phi^{\prime\prime}(r=\epsilon) = 0$ we can fix the integration
constant $c$ in (\ref{phi})
\begin{equation}
c^2 = 3\epsilon^2 ~.
\end{equation}
Finally the solution $\phi$ corresponding to a non-singular
transverse gravitational potential for the brane has the form
\begin{equation} \label{phi'}
\phi = \frac{3\epsilon^2+ ar^2}{3\epsilon^2+ r^2}~.
\end{equation}

From the condition that we have a 6-dimensional Minkowski metric
on the brane, $\lambda (r=0) = 1$, (any other value corresponds
only to a re-scaling of the extra coordinates)  we can fix also
the integration constant in (\ref{g})
\begin{equation}
\rho^2 = \frac{3\epsilon^2}{2(a-1)} ~.
\end{equation}
Then using (\ref{b}) the brane width can be expressed in terms of
the bulk cosmological constant and fundamental scale
\begin{equation} \label{epsilon}
\epsilon^2 = \frac{40M^4}{3\Lambda} ~.
\end{equation}
Now the metric tensor of the transverse space (\ref{g}) is not
dependent on $a$ and has the form
\begin{equation}\label{lambda}
\lambda = \frac{9\epsilon^4}{(3\epsilon^2 + r^2)^2} ~.
\end{equation}

Using solutions (\ref{phi}), (\ref{lambda}) and the relationship
(\ref{g}) to integrate the gravitational part of the action
integral (\ref{action}) over the extra coordinates we find
\begin{eqnarray}\label{R}
S_g= \frac{M^4}{2}\int dx^6\sqrt{-^6g}~^6R =
\frac{M^4}{2}\int_0^{2\pi} d\theta\int_0^\infty dr ~ r\phi^2\lambda
\int dx^4 \sqrt{-\eta}R \nonumber \\
= \rho^2 \pi M^4\int_1^a d\phi~\phi^2 \int dx^4 \sqrt{-\eta}R =
\frac{M^4}{2}\epsilon^2 \pi (a^2 + a + 1)\int dx^4 \sqrt{-\eta}R~,
\end{eqnarray}
where $R$ and $\eta$ are respectively the scalar curvature and
determinant, in four dimensions.

The formula for the effective Planck scale in our model, which is
two times the numerical factor in front of the last integral in
(\ref{R})
\begin{equation} \label{plank}
m_{Pl}^2 = M^4\pi\epsilon^2 (a^2 + a + 1) ~,
\end{equation}
is similar to those from the ``large" extra dimensions model
\cite{ADD}. The differences are, the presence of the value of
gravitational potential at extra infinity, $a$, in (\ref{plank}),
and that the radius of the extra dimensions is replaced by the
brane width $\epsilon$, which, as seen from (\ref{epsilon}), is
expressed by the ratio of the fundamental scale $M$ and the
cosmological constant $\Lambda$.

The normalization condition for a physical field, that its action
integral over the extra coordinates $r, \theta$ converges, is also
the condition for its localization. As was shown in \cite{GoSi}
Newtonian gravity is localized on the brane, since the action
integral for gravity, (\ref{R}), is convergent over the extra
space. However, the wave-functions of a localized matter field can
be spread out from the brane more widely then the brane width
$\epsilon$. In order not to have contradictions with experimental
facts, such as charge conservation \cite{DuRuTi}, the parameters
of the model must be chosen in a proper way.

When wave-functions of matter fields in six dimensions are peaked
near the brane in the transverse dimensions there wave-functions
on the brane can be factorized as
\begin{equation}\label{Xi}
\Xi (x^A) = \frac{\xi (x^\nu)}{\kappa } ~,
\end{equation}
where the parameter $\kappa$ is the value of the constant zero
mode with the dimension of length. These parameters can be found
from the normalization condition for zero modes
\begin{equation}\label{xi}
\int_0^{2\pi} d\theta \int_0^\infty dr
\sqrt{-^6g}\frac{1}{\kappa^2} = \sqrt{-\eta}~,
\end{equation}
which also guarantees the validity of the equivalence principle for
different kinds of particles.

Let us consider the situation with the localization of particular
matter fields. If we assume that the zero mode of a spin-0 field, $\Phi$, in six
dimensions is independent of the extra coordinates its action can be
brought to the form \cite{GoSi}
\begin{eqnarray} \label{action0}
S_\Phi = \int d^6x \sqrt{-^6g}~^6L_\Phi (x^A)
=\frac{2\pi}{\kappa^2_\Phi} \int_0^\infty dr ~ r\phi^2\lambda \int
d^4x \sqrt{-\eta}L_\Phi (x^\nu) =
 \nonumber \\
= \frac{2\pi\rho^2}{\kappa^2_\Phi} \int_1^a \phi^2d\phi \int d^4x
\sqrt{-\eta}L_\Phi (x^\nu) = \frac{\epsilon^2 \pi (a^2 + a +
1)}{{\kappa^2_\Phi}} \int d^4x \sqrt{-\eta}L_\Phi (x^\nu)~,
\end{eqnarray}
where $L_\Phi (x^\nu)$ is the ordinary 4-dimensional Lagrangian of
the spin-0 field and $\kappa_\Phi$ is value of the constant zero
mode. The integral over $r, \theta$ in (\ref{action0}) is finite
and the spin-0 field is localized on the brane.

The action for a vector field in the case of constant extra
components ($A_i = const $) also reduces to the 4-dimensional
Yang-Mills action multiplied an integral over the extra
coordinates \cite{GoSi}
\begin{eqnarray} \label{action1}
S_A = \int d^6x \sqrt{-^6g}~^6L_A (x^B)= \frac{2\pi}{\kappa^2_A}
\int _0^\infty dr ~ r\lambda \int d^4x \sqrt{-\eta}L_A (x^\nu) \nonumber \\
= \frac{2\pi\rho^2}{\kappa^2_A} \int_1^a d\phi \int d^4x
\sqrt{-\eta}L_A (x^\nu) = \frac{3\epsilon^2 \pi}{{\kappa^2_A}}
\int d^4x \sqrt{-\eta}L_A (x^\nu)~,
\end{eqnarray}
where $\kappa_A$ is the value of the zero mode of the vector
field. The extra integral in (\ref{action1}) is also finite and
the gauge field is localized on the brane.

The factorization of the zero mode of a 6-dimensional spinor field
in the ansatz (\ref{ansatzA}) is different from the definition
(\ref{Xi}), having instead the form \cite{GoSi}
\begin{equation} \label{B} \Psi (x^A) = \frac{\psi
(x^\nu)}{\kappa_\Psi \phi^2 \left( r \phi ' \right)^{1/4}}~ ,
\end{equation}
where $\kappa_\Psi$ is the value of the constant zero mode.
Integrating the 6-dimensional action of fermions over the extra
coordinates, using the explicit form (\ref{phi'}), yields
\begin{eqnarray} \label{action1/2}
S_\Psi = \int d^6 x \sqrt{-^6g}~^6L_\Psi (x^A)=
\frac{2\pi\rho^2}{\kappa^2_\Psi} \int_0^\infty dr~
\sqrt{\frac{\phi '}{r\phi^2}} \int d^4 x \sqrt{-\eta }L_\Psi
(x^\nu) \nonumber \\
= \frac{3\pi^2\epsilon^2}{\kappa^2_\Psi\sqrt{2a(a-1)}} \int d^4 x
\sqrt{-\eta }L_\Psi (x^\nu)~ ,
\end{eqnarray}
where $L_\Psi$ is the 4-dimensional Dirac Lagrangian. The extra
$1/\phi$ dependence in the second integral of (\ref{action1/2})
comes from the tetrad functions with upper index in the definition
of the Dirac gamma matrices for the ansatz (\ref{ansatzA}). The
integral in (\ref{action1/2}) over $r$ and $\theta$ is finite and
Dirac fermions are localized on the brane.

Equating the coefficients of action integrals (\ref{action0}),
(\ref{action1}) and (\ref{action1/2}) to $1$, so as to satisfy the
normalization condition (\ref{xi}), and to guarantee the equivalence
principle for gravity, we find the values of the zero modes for
spin $0$, spin $1$ and spin $1/2$ fields
\begin{equation} \label{xi1}
\kappa_\Phi^2 = \pi \epsilon^2 (a^2 + a + 1)~ , ~~~~~ \kappa^2_A =
3\pi \epsilon^2~ , ~~~~~ \kappa^2_\Psi =
\frac{3\pi^2\epsilon^2}{\sqrt{2a(a-1)}}~,
\end{equation}
which are used to parameterize the 4-dimensional
fields in the Lagrangians.

Within our model we now want to find the the positions where the
zero modes are localized as well as the localization radii of the
different fields.

From (\ref{action0}) the effective zero mode wave-function of the
scalar field in flat space can be defined as
\begin{equation} \label{zero0}
\Phi_0 (r)= \sqrt{\frac{2\pi r\phi^2\lambda}{\kappa^2_\Phi} }
= \frac{\sqrt{2\pi}3\epsilon^2}{\kappa_\Phi}
\sqrt{r}\frac{(3\epsilon^2+ ar^2)}{(3\epsilon^2+ r^2)^2}~,
\end{equation}
where $\kappa_\Phi$ has the value (\ref{xi1}). Function
(\ref{zero0}) is zero at infinity $(r\rightarrow \infty)$ and on
the brane $(r=0)$ and has a maximal value at some localization
distance, $d_\Phi$, between the brane ($r=0$) and infinity
($r=\infty$). This localization distance, $d_\Phi$, and the
localization radius $r_\Phi$, can be found by equating the first
and second derivatives (to find the maximum and inflection point
of the function) of (\ref{zero0}) to zero respectively. For the
localization distance this yields
\begin{equation} \label{rd0}
d_\Phi^2 = \epsilon^2 \frac{5a - 7 + \sqrt{49 -58 a +
25a^2}}{2a}~.
\end{equation}
From this formula we see that since $a>1$ the maximum of the
wave-function of scalar fields (\ref{zero0}) is located outside the
brane $d_\Phi > \epsilon$.

Setting the second derivative of (\ref{zero0}) to zero gives
\begin{equation} \label{rd0a}
5 a r^6 +(63 -66 a) \epsilon ^2 r^4 - (102 - 45 a) \epsilon ^4 r^2
- 9 \epsilon ^6 = 0~.
\end{equation}
This is an effectively cubic equation for $r^2$, which has one
real and two complex solutions. The radius, $r_{\Phi}$, of the
zero mode scalar wave-function is given by the real solution to
(\ref{rd0a}). The expressions of the solutions of (\ref{rd0a}),
which can be obtained using a symbolic mathematics program such as
{\it Mathematica}, are extremely long and we do not write them
here explicitly.

From (\ref{action1}) the effective wave-function of the vector
field zero mode takes the form
\begin{equation} \label{zero1}
A_0 (r) = \sqrt{\frac{2\pi r\lambda}{\kappa^2_A} } =
\frac{\sqrt{2\pi}3\epsilon^2}{\kappa_A}
\frac{\sqrt{r}}{3\epsilon^2+ r^2}~.
\end{equation}
This function is also zero on the brane and at the infinity, and
has a maximal value at some distance,  $d_A$, in between.
Again setting the first and second derivatives of
(\ref{zero1}) to zero we find
\begin{equation} \label{rd1}
d_A = \epsilon~, ~~~~~ r_A = \sqrt{ \frac{\epsilon ^2 (9 + 4
\sqrt{6})}{5}} \approx 1.9 \epsilon ~.
\end{equation}
So the peaks of the vector wave-functions are located exactly at the
edge of the brane, $r = \epsilon$ and the radius of localization is
approximately $2\epsilon$.

For the effective fermionic zero modes from (\ref{action1/2}) we
have
\begin{equation} \label{zero1/2}
\psi_0 (r) = \sqrt{\frac{2\pi\rho^2}{\kappa^2_\Psi} \left(
\frac{\phi '}{r\phi^2} \right)^{1/2}} =\left[\frac{54
\pi^2\epsilon^6}{\kappa_\Psi^4(a-1)}\right]^{1/4}
\frac{1}{\sqrt{3\epsilon^2+ ar^2}}~.
\end{equation}
This function is zero at infinity, but unlike the wave-functions
of the scalar and vector zero modes, the peak of the fermion
wave-function coincides with brane location, $r=0$. From the
inflection point of the function (\ref{zero1/2}), which is found
by equating the second derivative of (\ref{zero1/2}) to zero, we
obtain the localization radius for fermions
\begin{equation} \label{d1/2}
r_\Psi = \epsilon\sqrt{\frac{3}{2a}}  ~.
\end{equation}
From this formula we see that since $a > 1$ the peaks of fermionic
wave-functions on the brane are very sharp.

After introducing interactions between the various fields one
might be able to use the various overlappings of the wave
functions of the spin $0$, spin $1$ and spin $1/2$ fields in
speculations to explain the different types of mass hierarchies in
particle physics. For example, in the split fermion model
\cite{ArSc} localization of different species of fermions (not
fields with different spins as in our case) at different points of
a single thick brane was used to solve the hierarchy problem.

To summarize, in this paper it is shown that for a realistic form
of the brane stress-energy, there exists a static, non-singular
solution of the 6-dimensional Einstein equations, which provides
a gravitational trapping of 4-dimensional gravity and matter
fields on the brane. An essential feature of the model is that
different kinds of matter fields have different localization
distances from the brane. This property is in principle
experimentally testable.

\begin{flushleft}
{\bf Acknowledgments} This work is supported by a 2003 COBASE
grant.
\end{flushleft}

\end{document}